\documentstyle[psfig,amstext,12pt]{article}

\title{Warped Geometry of Brane Worlds}
\author{Gary N. Felder, Andrei Frolov, and Lev Kofman\\
{\small CITA, University of Toronto,
60 St. George St., Toronto ON  Canada, M5S 3H8}}
\date{18 December 2001}

\newcommand{\picdir}[1]{./#1}
\newcommand{\K}{{\cal K}}
\renewcommand{\H}{{\cal H}}

\begin{document}
\baselineskip=14.5pt
\pagestyle{plain}
\setcounter{page}{1}

\begin{titlepage}

\maketitle

\begin{abstract}
We study the dynamical equations for extra-dimensional dependence of a
warp factor and a bulk scalar in 5d brane world scenarios with induced
brane metric of constant curvature. These equations are similar to
those for the time dependence of the scale factor and a scalar field in
4d cosmology, but with the sign of the scalar field potential reversed.
Based on this analogy, we introduce novel methods for studying the
warped geometry. We construct the full phase portraits of the warp
factor/scalar system for several examples of the bulk potential. This
allows us to view the global properties of the warped geometry. For
flat branes, the phase portrait is two dimensional. Moving along
typical phase trajectories, the warp factor is initially increasing and
finally decreasing. All trajectories have timelike gradient-dominated
singularities at one or both of their ends, which are reachable in a
finite distance and must be screened by the branes. For curved branes,
the phase portrait is three dimensional. However, as the warp factor
increases the phase trajectories tend towards the two dimensional
surface corresponding to flat branes. We discuss this property as a
mechanism that may stretch the curved brane to be almost flat, with a
small cosmological constant. Finally, we  describe the embedding of
branes in the 5d bulk using the phase space geometric methods developed
here. In this language the boundary conditions at the branes can be
described as a 1d curve in the phase space. We discuss the naturalness
of tuning the brane potential to stabilize the brane world system.

\bigskip
\noindent
CITA-2001-82
\end{abstract}

\thispagestyle{empty}
\end{titlepage}

\section{Introduction}

One of the most interesting recent directions in high energy physics
phenomenology is the development of brane world scenarios in which our
3+1 dimensional spacetime is a 3-brane embedded in a higher dimensional
spacetime. In application to the very early universe this leads to
brane world cosmology, where the universe we observe is a 3+1
dimensional curved brane embedded in the bulk. One of the issues in
brane world scenarios is the warped geometry of the internal space. In
addition to the warp factor in the bulk, brane world scenarios often
contain bulk scalar fields. Examples include the dilaton in
Horava-Witten theory \cite{HW} (associated with the volume of the
compactified 6d Calabi-Yau space) where  the 5d effective theory can be
obtained \cite{Lukas}; the Randall-Sundrum model \cite{RS1} with
phenomenological stabilization \cite{GW} where the choice of the 
bulk/brane potentials  must be consistent with the 5d warp geometry
\cite{Dewolfe,GKL}, the scalar sector of the supergravity realization
of the Randall-Sundrum model \cite{AVP}, bulk supergravity with domain
walls \cite{ST} and others.

The 5d bulk scalar plus gravity brane world system is based on the five
dimensional Einstein equations with junction conditions at the branes.
Here we consider a simpler problem where the 5d space can be split
into 4+1 de~Sitter slices
\begin{equation} \label{warp}
ds^2 = dw^2+ A^2(w)ds_4^2 \, .
\end{equation}
The 4d de~Sitter geometry is described by its scalar curvature
${^4}R=12H^2$, where $H=\text{const}$ is the 4d Hubble parameter. The
warp factor $A(w)$ is determined up to boundary conditions by the five
dimensional Einstein equations. For the sake of generality we also
present corresponding results for the more general case of a $D$
dimensional warped metric with $D-1$ dimensional de~Sitter slices. The
limit of vanishing $H$ corresponds to a flat brane, while nonvanishing
$H$ corresponds to brane inflation.

The brane sets the boundary conditions for the warp factor $A(w)$ and
the scalar field $\phi(w)$. It is known that often the warped geometry
(\ref{warp}) somewhere outside of the brane encounters a spacetime
singularity. One way to cure this problem is to invoke a second brane
to screen the singularity  by making the inner geometry periodic with
the inter-brane interval. Two end-of-the-world branes provide orbifold
compactification of the inner space. In the Randall-Sundrum model
\cite{RS1} with AdS bulk geometry without scalars the second brane may
be removed. 

The properties of warped geometry with one or two branes were studied
in many papers, see e.g.~\cite{CR,ST,Dewolfe,GKL,FTW,Davis}. The
purpose of this paper is to investigate the global properties of  5d
warped geometry (\ref{warp}) for a variety of bulk scalar field
potentials $V(\phi)$, supplemented by boundary conditions at the
branes. We will try to understand how typical is the singularity in the
warped geometry, how much tuning is required for the brane potentials,
and how these depend on the brane curvature $H^2$. Our approach is
different from what was used in the earlier literature.

The setting of the problem for the geometry (\ref{warp}) is similar to
the investigation of the FRW universe geometry
\begin{equation} \label{fr}
ds^2 = -dt^2+ a^2(t)ds_3^2 \, .
\end{equation}
with the scale factor $a(t)$ and a scalar field with the potential
$V(\phi)$.
Powerful method that has been used to investigate this 4d problem is
the construction of phase portraits for the dynamic system for
variables $\phi$, $\dot{\phi}$, and $\dot{a}/a$. Using this method it
can be shown that for a broad range of potentials $V(\phi)$ inflation
occurs along a separatrix that is a typical intermediate asymptotic for
a broad band of phase trajectories \cite{acad,KLS}.

Inspired by this analogy, we adopt the phase portrait approach to
studying the warped geometry of the brane world scenario. It turns out
that the equations for the system $(A(w), \phi(w))$ with the potential
$V(\phi)$ are similar to the cosmological equations for $(a(t),
\phi(t))$ but with the sign of the potential reversed. (There are also
differences in the numerical coefficients in 4d and 5d.)  Flipping the
sign of the potential makes a big difference.  For example, it alters
the geometry of the phase portrait by connecting branches with positive
and negative ``Hubble'' parameter ${A' \over A}$.  This connection with
4d cosmology suggests a convergence of this work with recent work on 4d
cosmology with negative potentials \cite{fastroll}, the results of which can
be extended to the warped geometry. Our results overlap with
\cite{fastroll} and the connection will be investigated further
\cite{FFKL}.

The structure of the paper is the following. In Section 2, we introduce
the basic equations for the brane world scenario. In Section 3, we
discuss generic properties of the brane world phase space in terms of
${A' \over A}, \phi', \phi$ and classify its critical points.

In Section 4, we systematically construct the phase portrait for a 5d
space with flat 4d curvature $H=0$ (flat branes) for the simple
quadratic potential $V(\phi)={1 \over 2}m^2 \phi^2$.  We will see that
without branes all trajectories begin and end at naked singularities
dominated by the gradient energy $\phi'^2$ of the scalar field, which
corresponds to a ``stiff'' equation of state with anisotropic pressure.
We also consider quadratic potentials  with positive and negative
cosmological constants.

In Section 5, we consider exponential potentials
$V(\phi)=V_0 e^{-2\sqrt{2}\phi}$.

In Section 6, we extend the method of phase portraits to brane world
scenarios with curved branes $H \ne 0$. We shall see that the brane
with the larger warp factor will have smaller curvature. We discuss how
this effect may be related to the problem of the small cosmological
constant on the visible brane.

In Section 7, we derive the Hamilton-Jacobi form of the self-consistent
Einstein equations for warped geometry with a scalar field, which leads
to the SUSY form of an arbitrary positive bulk scalar potential
(without any underlying supersymmetry). This correspondence has been
previously noted in context of holographic renormalization group flows
\cite{deBoer:1999}. We also address the similarity of the
Einstein-Hamilton-Jacobi constraint equation and the well-known
gravitational stability form of the potential \cite{susy1,susy2,ST}.

In Section 8, we introduce branes to screen the singularities. We show
how the brane boundary conditions can be represented geometrically as a
1d curve in the 3d phase space of the system.  It turns out to be
convenient to use the EHJ formalism (in many respects similar to using
the SUSY form of the potential). From this perspective we will discuss
potentials that lead to brane stabilization and the degree of
fine-tuning required to achieve them.

The paper concludes with the summary of our results. There is also an
appendix in which the locations of critical points at infinity         
are derived for the phase portraits shown here.

\section{Equations and Notation}

In this section we give the general formalism for a brane world
scenario with a bulk warp factor and scalar field. The total action is
\begin{eqnarray}\label{action}
S &=& \frac{1}{16\pi \kappa_D^2} \int \sqrt{-g}\, d^D x\,
           \left\{R - (\nabla\phi)^2 - 2V(\phi)\right\} \\
  && -\frac{1}{8\pi \kappa_D^2} \sum \int \sqrt{-h}\, d^{D-1} x\,
           \left\{ [\K] + U(\phi)\right\}. \nonumber
\end{eqnarray}
The $D$-dimensional gravitational coupling is related to the
$D$-dimensional Planck mass by $M_D^{D-2}=\frac{1}{8\pi
\kappa_D^2}$. The bulk scalar field $\phi$ in (\ref{action}) is
dimensionless. The physical value of the scalar field with canonical
normalization is $\Phi={\phi \over { \sqrt{8\pi}\kappa_D}}$ and the
physical scalar field potential is ${ V \over {8\pi \kappa_D^2}}$.
The first term in (\ref{action}) describes the bulk; the second term
is related to the brane(s).  We use a ``mostly positive'' signature
and the curvature conventions of Misner, Thorne and Wheeler (MTW). We
write the jump of a quantity across the brane as $[\K ]={\K}^+-{\K}^-$.
Bulk indices will be greek, $\mu, \nu, ...$; brane indices will be
latin, $a, b, ...$. Brane hypersurfaces are denoted as $\Sigma_i$ where
the index $i$ runs over all the branes. Throughout the paper we use
overdots to indicate time derivatives and primes to indicate
derivatives with respect to the fifth dimension $w$.

The brane extrinsic curvature $\K_{ab}$ is expressed through the
normal unit vector $n_\mu$ and the tangent vierbien $e_{(a)}^\mu$
\begin{equation}\label{extr}
\K_{ab} = e_{(a)}^\mu e_{(b)}^\nu \nabla_\mu n_\nu \ ,
\end{equation}
the bulk equations are
\begin{equation}\label{bulk}
R_{\mu\nu}-{ 1 \over 2} R g_{\mu\nu} = T_{\mu\nu}, \hspace{1em}
\Box\phi = \frac{\partial V}{\partial\phi},
\end{equation}
and the scalar field stress-energy tensor is given by
\begin{equation}\label{em}
T_{\mu\nu} = \phi_{,\mu} \phi_{,\nu} + \left(-\frac{1}{2}
(\nabla\phi)^2 - V(\phi)\right) g_{\mu\nu}.
\end{equation}

The  junction conditions  are
\begin{equation}\label{junction}
[\K_{ab} - \K g_{ab}] = U(\phi) g_{ab}, \hspace{1em} [n \cdot
\nabla\phi] = \frac{\partial U}{\partial\phi}.
\end{equation}

For some brane world scenarios the bulk and brane scalar field
potentials are known. In case of the RS models, for example, we have
$U(\phi)=\pm \lambda$, where $\lambda$ is a constant with different
signs on the two branes. The bulk potential is just a 5d negative
cosmological constant $V(\phi)=-\Lambda$. In case of the HW model,
$U(\phi)=U_0 e^{-\sqrt{2} \phi}$, $V(\phi)=V_0 e^{-2 \sqrt{2}
\phi}$. Our calculations will be valid for arbitrary $U(\phi)$ and
$V(\phi)$.

For 5d warped geometry (\ref{warp}) the bulk Einstein equations can be
written as
\begin{equation}\label{einstein1}
\left({A' \over A}\right)' =-{1 \over 4} \phi'^2 -{1\over 6} V(\phi) -
\left({A' \over A}\right)^2,
\end{equation}
\begin{equation}\label{einstein2}
6 \left({A' \over A}\right)^2={1 \over 2} \phi'^2- V(\phi)+6{ H^2
\over A^2} \ ,
\end{equation}
where the latter is a constraint equation. The equation for bulk
scalar field
\begin{equation}\label{phiequation}
\phi'' + 4 {A' \over A} \phi' - V_{,\phi} = 0
\end{equation}
is redundant with the Einstein equations.

In addition to the bulk equations given above we can specify boundary
conditions at the brane.  The assumption of $Z_2$ symmetry around each
brane $\Sigma$ implies the boundary condition for the warp factor
\begin{equation}\label{junction1}
{A' \over A} \vert_{\Sigma}= -{1 \over 6} U
\end{equation}
and for the scalar field
\begin{equation}\label{junction2}
\phi'  \vert_{\Sigma}=   {1 \over 2} U_{,\phi}.
\end{equation}
Unless otherwise specified all boundary conditions are given on the
positive side of the brane.

When we refer to a {\it flat} or {\it curved} brane we are referring
to its 4d, i.e. spacetime, curvature. We only consider spatially flat
branes, so the term {\it flat brane} refers to a 3+1 Minkowski
geometry while a {\it curved brane} refers to a 3+1 de~Sitter
geometry.

The Einstein and field equations (\ref{einstein1}-\ref{phiequation})
can in principle be solved to find the $w$ dependence of $\phi$ and
$A$, with boundary conditions supplied by the junction conditions
(\ref{junction1}-\ref{junction2}).  It is instructive to compare 5d
warped geometry with a scalar field with 4d cosmological geometry
(\ref{fr}) with a scalar field.  In 4d cosmology we have the Einstein
equations
\begin{equation}
\label{einstein3}   \left({\dot a  \over a}\right)^.
 =- { 1 \over 3}\dot \phi^2 +{ 1 \over 3} V(\phi)
 -  \left({\dot a  \over a}\right)^2 \
\end{equation}
\begin{equation}\label{f}
3\left({\dot a  \over a}\right)^2=
 {1 \over 2}\dot \phi^2+V(\phi) - 3 {K\over a^2}  \ ,
\end{equation}
where $K=0, \pm1$ is the curvature of 3d space,
plus a redundant scalar field equation
\begin{equation}\label{phiequation1}
\ddot \phi + 3 { \dot a \over a} \dot \phi + V_{,\phi}=0  \ .
\end{equation}

Apart from trivial numerical coefficients, equations
(\ref{einstein1}-\ref{phiequation}) and
(\ref{einstein3}-\ref{phiequation1}) differ in only one respect:
for the same sign of the potential in the action (\ref{action}) the
sign of the potential in the cosmological equations will be opposite
for the 4d cosmology and 5d brane world cases.  The sign
reversal comes about because of the different metric signatures of
$dw^2$ and $dt^2$.

The 4d cosmological equations for positive potentials have been
comprehensively studied by means of qualitative methods for analyzing
ODEs \cite{KLS,acad}.  These methods are equally applicable to 5d
warped geometry.  However, we will see below that the sign reversal
has profound implications for the qualitative properties of the phase
space behaviour.

Throughout the paper we will switch back and forth between a 4d and 5d
viewpoint for the potentials we consider. We will usually talk about
the shape of our potentials in terms of the behavior they elicit. For
example, we refer to the potential $V=\frac{1}{2} m^2 \phi^2$ in 5d warped geometry
 (\ref{phiequation}) as a hill rather than a well, reflecting
the fact that $\phi$ will tend to accelerate away from the origin
rather than towards it for this case. The same potential acts as a well
in the context of the 4d cosmological equation (\ref{phiequation1}). 
We will examine the implications of our work for 4D cosmology with
negative potentials in greater depth in a subsequent publication
\cite{FFKL}.

\section{Phase Portraits}

Dynamical system  (\ref{einstein1}-\ref{phiequation}) can be rewritten in
terms of the variables
\begin{equation}
x \equiv \phi;\;y \equiv \phi';\;z \equiv {A' \over A},
\end{equation}
which gives the three ``evolution'' equations
\begin{eqnarray}
\label{phaseequation1}x' &=& y \\
\label{phaseequation2}y' &=& V_{,x} - 4 y z \\
\label{phaseequation3}z' &=& -{1 \over 6} V - {1 \over 4} y^2 - z^2,
\end{eqnarray}
plus the constraint equation
\begin{equation}\label{constraint}
-2 V + y^2 - 12 z^2 = -12 {H^2 \over A^2} \  .
\end{equation}
Here $V=V(x)$.

Reduction of equations to a set of first order ODEs allows us to
represent their solutions using phase portraits, i.e. plots of
trajectories in the 3d phase space defined by $x$, $y$, and $z$. This
technique has been applied to the evolution equations for 4d chaotic
inflation models in \cite{acad,KLS}. In general, all trajectories in
phase space must begin and end at critical points, i.e. points where
the derivatives of all the dynamic variables, i.e. the {\it r.h.s.}
of (\ref{phaseequation1}-\ref{phaseequation3}), vanish. In addition to
starting and ending at such points, trajectories may in general come
from or move towards infinity. By making a coordinate transformation
of the phase space coordinates $(x,y,z)$ that projects the complete
phase space onto a compact region (Poincar\'{e} projection),
 however, one can define a discrete
set of critical points at infinity. Together with the set of finite
critical points these points describe the complete set of possible
beginning and ending points for all trajectories. By identifying the
properties of these critical points (attractors $A$, repulsors $R$ or
saddle points $S$) and the trajectories that connect them it is
possible to obtain a complete qualitative description of the dynamical
system (\ref{phaseequation1}-\ref{phaseequation3}).
(We do not expect more complicated situations for our dynamical system.)

Note that the parameter $H$ related to the curvature of the branes
appears only in the constraint equation (\ref{constraint}). Just as for
standard FRW cosmology,
  our equations for warped bulk geometry can be classified
by the sign of $H^2$. The case $H^2>0$ corresponds to curved branes
with 4d de~Sitter spacetime geometry, $H^2 < 0$ corresponds to 4d AdS
space-time, and $H=0$ corresponds to flat branes. These three cases
correspond to three regions in phase space. Since $H^2$ is a constant
of the motion, the phase space trajectories can never cross from one
of these three regions to another. In particular this means that for
any particular potential $V$ the surface obtained by setting $H^2=0$
in equation (\ref{constraint}) defines a limiting surface in phase
space that can never be crossed. In later sections we will see the
importance of this two dimensional surface for defining the phase
space portrait for different potentials. In this paper we only
consider trajectories in the regions with $H^2 \geq 0$.

In the rest of this section we discuss finding the critical points for
a general potential $V(x)$. In subsequent sections we illustrate this
general procedure with a series of examples for which we construct the
phase portraits.

To find the finite critical points we set all {\it r.h.s.} of
equations (\ref{phaseequation1}-\ref{phaseequation3}) to zero. This
gives rise to the conditions
\begin{equation}
y = V_{,x} = 0,
\end{equation}
\begin{equation}
z^2 = -{1 \over 6} V(x) \ .
\end{equation}
For positive definite potentials there can be no finite critical
points.  For instance, for $V(\phi)=V_0 e^{-2\sqrt{2}\phi}$
 there are no finite
critical points.  For non-negative potentials finite critical points
must all satisfy the conditions
\begin{equation}
y = z = V = V_{,x} = 0.
\end{equation}
Any extremum for which $V<0$, however, corresponds to two critical points
\begin{equation}
z = \pm \sqrt{-{1 \over 6} V}.
\end{equation}

To analyze the behavior near these critical points $(x_0, y_0, z_0)$
we consider small deviations
\begin{equation}
x = x_0 + \delta x;\;y = \delta y;\;z = z_0 + \delta z
\end{equation}
and linearize the equations
(\ref{phaseequation1}-\ref{phaseequation3}). We then assume a solution
of the form $(\delta x, \delta y, \delta z) \sim e^{\lambda w}$, which
gives us a matrix with eigenvalues
\begin{equation}
\lambda = \left(-2 z_0 \ , \,\,\, -2 z_0 \pm \sqrt{V_{0,xx} + 4 z_0^2}
 \right) \ .
\end{equation}
Negative eigenvalues correspond to attractors and positive eigenvalues
to repulsors. Critical points whose eigenvalues have different signs
are saddle points. Those with imaginary eigenvalues show oscillatory
behavior.  In
our case the finite critical points can not be stable for $z_0<0$ or
for $V_{0,xx}>0$. In either of these cases there will be at least one
unstable direction in the vicinity of the critical point. For $z_0>0$
and $V_{0,xx}<0$ the solutions are stable.

It still remains to find critical points that occur at infinite values
of the parameters. To do this we rescale the infinite space of $x$,
$y$, and $z$ into a finite Poincar\'{e} sphere by means of the variable
definitions
\begin{eqnarray}
x &=& {r \over 1-r} \cos(\theta) \sin(\varphi) \\
y &=& {r \over 1-r} \sin(\theta) \sin(\varphi) \\
z &=& {r \over 1-r} \cos(\varphi).
\end{eqnarray}
We also shall rescale $w$ by defining $d\tilde{w} = dw/(1-r)$.

Our phase space as described by the spherical coordinates $\{r,
\varphi, \theta\}$ is contained within a sphere of radius one. The ODEs
describing our system in these variables are written in Appendix~A.
Here we are only concerned with critical points at infinity, which
corresponds to $r=1$. In general the terms involving $V(x)$ and
$V_{,x}(x)$ may be divergent for large $x$, in which case they alter
the structure of the infinite critical points. Assuming $V$ and
$V_{,x}$ are not divergent as $x \to \infty$, however, the set of
infinite critical points turns out to be independent of the potential.
Given this assumption, we found 8 infinite critical points. We label
them $(S_1,S_2,S_3,S_4,A_1,A_2,R_1,R_2)$, reflecting their behavior for
the potentials we have examined. ($S \leftrightarrow$ saddle point, $A
\leftrightarrow$ attractor, $R \leftrightarrow$ repulsor.) Their
coordinates on the Poincar\'{e} sphere are at $r=1$ and
\begin{eqnarray}\label{infcrit}
(\varphi,\theta)
&=&(0,0), ~ (\pi/2,0), ~ (\pi/2,\pi), ~ (\pi,0), ~ \\
&& (\sin^{-1}(\sqrt{12/13}) \ , \pi/2), ~
   (\sin^{-1}(\sqrt{12/13}), 3 \pi/2), ~ \nonumber\\
&& (\pi-\sin^{-1}(\sqrt{12/13}), \pi/2), ~
   (\pi-\sin^{-1}(\sqrt{12/13}), 3 \pi/2). \nonumber
\end{eqnarray}
For the potentials considered in this paper several  attractor and repulsor
points are  located on the two dimensional surface
corresponding to $H^2=0$.

\begin{figure}
\centerline{\psfig{figure=\picdir{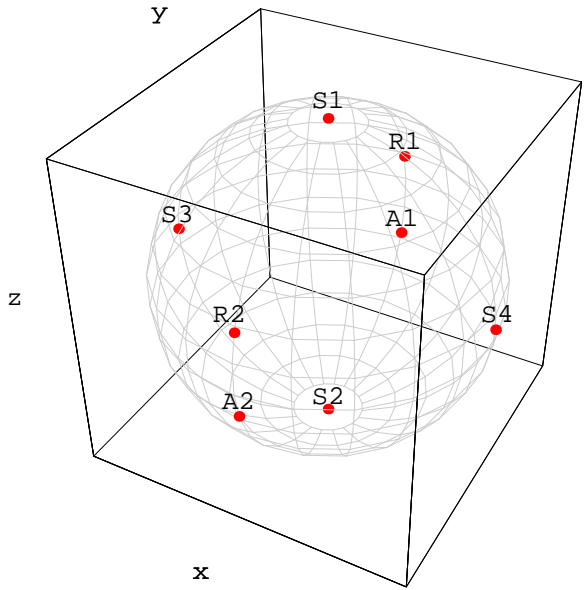},width=.5\columnwidth}}
\caption{Infinite critical points on the Poincar\'{e} sphere. The labels
are given to show correspondences with later plots for specific
potentials.}\label{fig:crit}
\end{figure}

Most realistic potentials are divergent as $x \to \infty$, with the
result that infinite critical points can be added or removed relative
to this generic picture. For instance, for $V={1 \over 2} m^2 \phi^2$,
$V=-{1 \over 2} m^2 \phi^2$, and $V=V_0 e^{-\phi}$ there are six,
twelve, and seven infinite critical points respectively. Nonetheless a
close correspondence can usually be seen between the critical points
for these potentials and the generic ones discussed here.

\section{Quadratic Bulk Potentials: $V = \frac{1}{2}\, m^2 \phi^2$}\label{sec:quadratic}

As a simple example of a bulk scalar potential we consider $V = \frac{1}{2}
m^2 \phi^2$. In the equations of motion the mass $m$ can be absorbed
by rescaling the fifth coordinate $w \to m w$, so without loss of
generality we simply set $m=1$. Thus in the general equations
(\ref{phaseequation1}-\ref{phaseequation3}) we shall use
\begin{equation}\label{quadratic}
V = {1 \over 2} x^2, \hspace{1em}
V_{,x} = x.
\end{equation}
There is one finite critical saddle point for this case at the origin
$x=y=z=0$. In the field equation (\ref{phiequation1}) this point
corresponds to the field sitting at the top of the potential with no
velocity and the warp factor  neither increasing nor decreasing. This
state is, however, unstable. As we show in the Appendix, there are
only six infinite critical points for this case at
\begin{eqnarray}\label{ip}
(\varphi,\theta)
&=& (0,0), ~ (\pi,0), ~ (\sin^{-1}(\sqrt{12/13}), \pi/2), ~
(\sin^{-1}(\sqrt{12/13}), 3 \pi/2), \\ \nonumber &&
(\pi-\sin^{-1}(\sqrt{12/13}), \pi/2), ~ (\pi-\sin^{-1}(\sqrt{12/13}), 3
\pi/2).
\end{eqnarray}
In other words the set of infinite critical points is identical to the
generic case (\ref{infcrit}) except without the points $(\pi/2,0)$ and
$(\pi/2,\pi)$, i.e. $S3$ and $S4$ in Figure~\ref{fig:crit}.

\begin{figure}
\centerline{\psfig{figure=\picdir{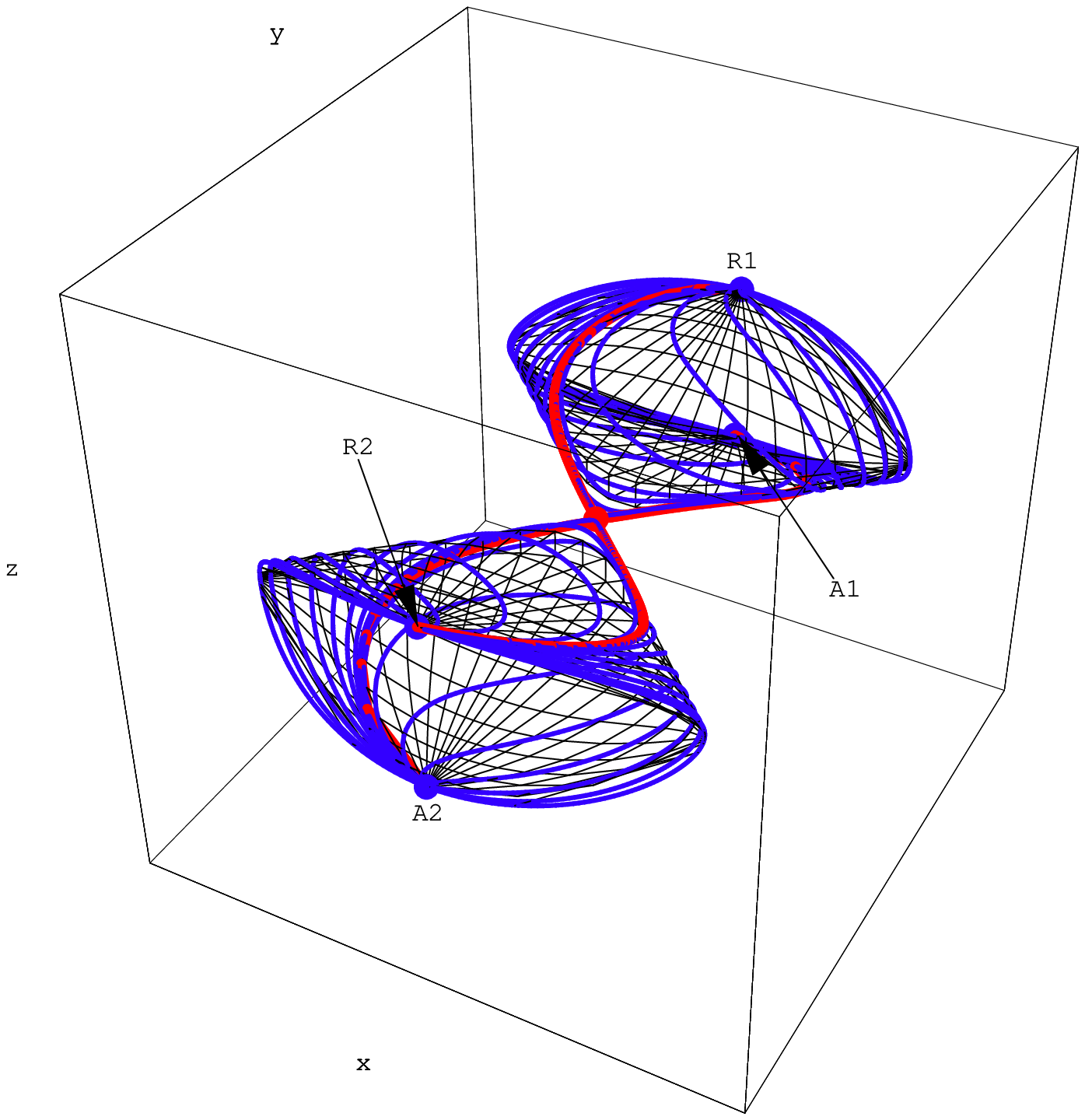},width=.75\columnwidth}}

\begin{minipage}{.45\columnwidth}
\centerline{\psfig{figure=\picdir{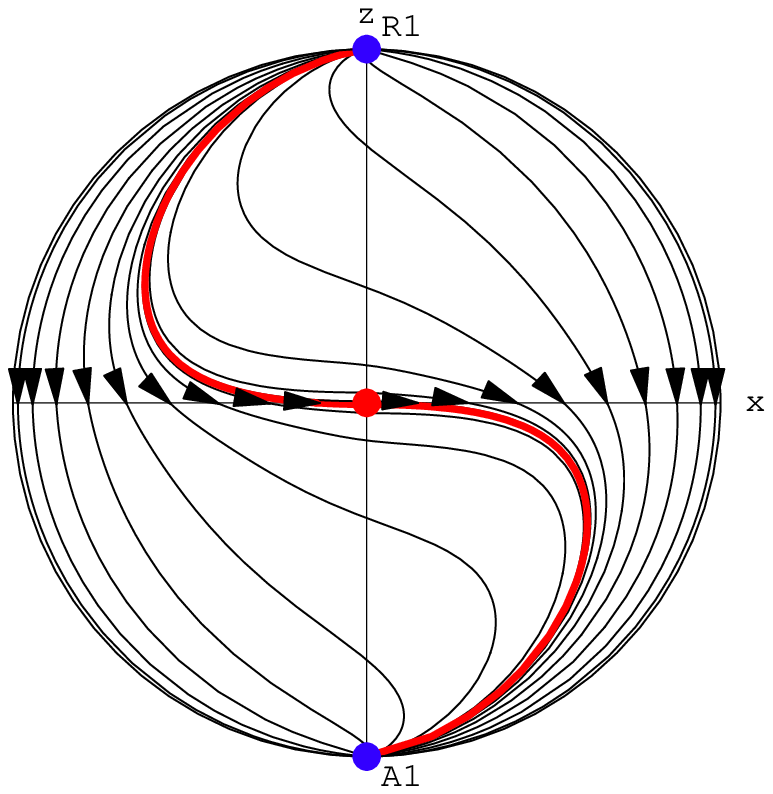},width=\columnwidth}}
\end{minipage}
\hfill
\begin{minipage}{.45\columnwidth}
\centerline{\psfig{figure=\picdir{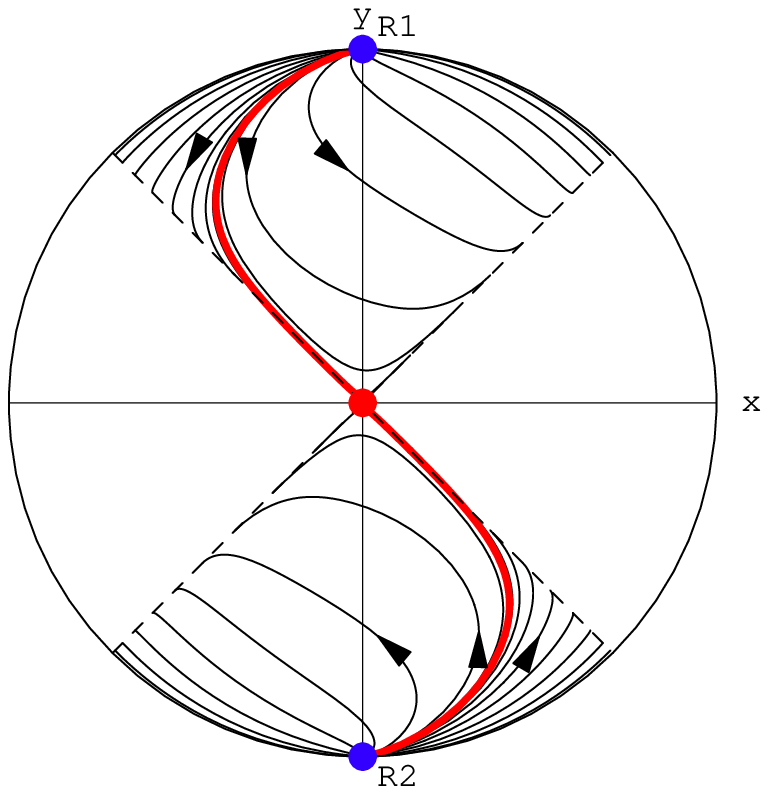},width=\columnwidth}}
\end{minipage}
\caption{ Phase portrait in Poincar\'{e} coordinates and 
trajectories with $H^2=0$ for $V=\frac{1}{2} m^2 \phi^2$. The top
figure shows the trajectories in the full 3D phase space while the
bottom pictures show projections onto the $xz$ and $xy$ planes. In
both projections, only half of the cone is shown for clarity.
Trajectories at $xy$ plane continue on the other side of the cone.}
\label{m2x2h0}
\end{figure}

The constraint equation (\ref{constraint}) for the quadratic potential
has the form
\begin{equation}\label{mx2constr}
-x^2+ y^2 - 12 z^2 = - 12 { H^2 \over A^2} \ .
\end{equation}
Setting $H^2=0$ in the constraint equation defines the two dimensional
surface
\begin{equation}\label{m2x2constraint}
-x^2 + y^2 - 12 z^2 = 0,
\end{equation}
which is a double cone opening up in the positive and negative $y$
directions, see upper panel of Figure~\ref{m2x2h0}.

The cases $H^2>0$ and $H^2<0$ correspond to trajectories outside and
inside the double cone respectively. For all values of $H^2$ nearly all
trajectories begin at the points $R_1$ and $R_2$ on the top of the cone
and end at the points $A_1$ and $A_2$ on the bottom, see
Figures~\ref{m2x2h0} and \ref{m2x2}. There are 
separatrices that connect saddle infinite points $S_1$, $S_2$ with 
the saddle point $S$ at the origin, see Figure~\ref{m2x2}.

In this section we consider trajectories that lie along the cone,
i.e. ``flat'' geometry with $H^2=0$. As we have
mentioned, this case plays a special role in the investigation of the
3d phase portrait. We consider the case $H^2>0$ in section
\ref{dsspace}. Figure~\ref{m2x2h0} shows
Poincar\'{e} projection of two dimensional phase space and 
 some of the trajectories for
the flat case. Phase trajectories that begin on the double cone
surface (\ref{m2x2constraint}) remain on it. Trajectories that lie
off the cone can, and in general do, approach it in the limit $r \to
1$, but any trajectory that lies on the cone at any finite point must
lie entirely on it. Therefore, the phase space required to describe
warped bulk geometry with flat 4d slices is two dimensional.

The flat case we are considering is similar to 4d FRW flat cosmology
with a scalar field, which can also be described with a two
dimensional phase portrait \cite{acad, KLS}. In the cosmological
case the phase manifold has two disconnected planes, one for an
expanding universe and another for a contracting universe (which can
be obtained from the first by time reversal). In the 5d case, however,
the expanding and contracting regions of the phase portrait are
connected, which leads to very different behavior of $A(w)$ and
$\phi(w)$.

In addition to showing the 3d phase portrait, Figure~\ref{m2x2h0} also
shows the projection of one of the cones onto the $(x, z)$ plane and
the projection of the upper half of both cones onto the $(x, y)$
plane. There are several interesting features to note about this
portrait. As in the 4d case, trajectories can not pass from one cone
to the other. Trajectories can not pass through the critical point $S$
at the origin, or through any critical point for that matter, because
by definition all derivatives vanish at these points. In the 4d case
this separation meant that for $H^2=0$ with this scalar
potential the sign of the Hubble parameter could never change. In the
5d case, by contrast, the geometry of the cone dictates that the sign
of $y$, i.e. of $\phi'$, can never change. Consider what this means
for the equation of motion (\ref{phiequation}), which describes an
inverted harmonic potential with friction
\begin{equation}\label{osc}
x''+4{A' \over A}x' -x=0 \ .
\end{equation}
Ordinarily we would expect that if you start on one side of the hill
$-x^2$ rolling towards the origin you could either go over the top or
stop and roll back in the direction you came from. However, the latter
behavior is not possible for our dynamical
 system.  The reason comes from the
behavior of the effective ``friction'' ${A' \over A}=z$. Looking at the
constraint equation (\ref{m2x2constraint}) we can see that as $y$
decreases $z^2$ approaches zero. Once the kinetic energy is small
enough $z$ will change sign. This means that the warp factor $A(w)$,
which was initially increasing, will begin shrinking.
 At this turn-around point the friction term
in the scalar field equation of motion (\ref{osc}) becomes negative,
$y$ begins to grow again, and the field inevitably makes it over the
top of the hill. From that point onward $y$ grows with increasing
speed, eventually reaching the singular point at infinity within a
finite distance. (If we insert brane at some $w_1$ so that
 $A(w)$ was decreasing initially then friction
will always be negative, and the same result holds.)

This behavior is clear in the phase portraits shown here. All
trajectories  must cross from $z>0$ to $z<0$, and must end up at the
infinite critical point with ${x \over y} \to 0$. In other words as
$\phi$ grows, $\phi'$ grows infinitely faster, so the singularity is
reached in a finite distance. Note that on each branch of the cone
($\pm y$) there are two topologically distinct sets of trajectories,
distinguished by which side of the cone ($\pm x$) they are on when
they cross from $z>0$ to $z<0$. Physically this difference simply
reflects whether they crest the top of the hill before or after the
turn-around point discussed earlier. These two classes of trajectories
are separated by separatrices connecting the infinite critical points
to the origin, as can be seen in the right-hand side of Figure~\ref{m2x2h0}.

All trajectories begin on the repulsive infinite
critical points $R_1$ and $R_2$ where $z >0$ and end at the attractive
infinite critical points $A_1$ and $A_2$ where $z < 0$. This is why
eventually $A(w)$ always decreases.  The  separatrix
trajectories are obtained from two curves that intersect at the origin,
one each starting from $R_1$ and $R_2$ and ending at $A_2$ and $A_1$
respectively.

Near the end points of all  trajectories $R_1$, $R_2$, $A_1$, and $A_2$
the gradient term ${1 \over 2} y^2$ dominates over the potential term
${1 \over 2} x^2$. Recall that energy density of the the scalar field
from (\ref{em}) is $\rho = {1 \over 2}\phi'^2+{1 \over 2}m^2 \phi^2$;
its pressure is anisotropic: $P_w={1 \over 2}\phi'^2-{1 \over 2}m^2
\phi^2$, while in the other three directions $P_{1,2,3} = -{1 \over
2}\phi'^2-{1 \over 2}m^2 \phi^2$. Thus, in the regime where gradient
terms dominate, we have $T^{\mu}_{\nu}=-\rho \,\, \text{diag}(+1, +1,
+1, +1, -1)$. (Recall that we use MTW convention for $T^{\mu}_{\nu}$
where $\rho=-T^{0}_{0}$). The pressure is anisotropic, in the direction
of the extra dimension it corresponds to a stiff equation of state $P_w
=\rho$, while in the perpendicular direction it has vacuum-like
equation of state $P_{1,2,3}=-\rho$. Compare this with a stiff equation
of state in 4d scalar field cosmology nearby spacelike singularity,
i.e.~$P =\rho$ \cite{acad,KLS}. As we approach infinite values of
$\phi'$, we have $\rho \propto A^{-2(D-1)}$, in 5d $\rho \propto
A^{-8}$. This corresponds to a timelike singularity (say at $w=w_0$) as
$A(w) \to 0$. Near singularities $\phi \propto \sqrt{\frac{D-2}{D-1}} \log
(w-w_0)$, $A(w) \propto (w-w_0)^{\frac{1}{D-1}}$. In 5d, $A(w) \propto
(w-w_0)^{1/4}$. We conclude that the end points of all trajectories
correspond to timelike singularities. For realistic models these
singularities should be screened by branes. 

To end this section we modify the quadratic potential by adding a
cosmological constant $V = 1/2 m^2 \phi^2 + \Lambda$, where $\Lambda$
may have either sign. Negative $\Lambda$ with no scalar (or
equivalently $m=0$) corresponds to the Randall-Sundrum \cite{RS1,RS2}
model. In other words we are considering the question of what happens
to the RS model (say AdS with a single flat brane) if one adds a
massive bulk scalar. As we will see, this model has the same gradient
type naked singularities that we found for the simple quadratic
potential.

\begin{figure}
\centerline{\psfig{figure=\picdir{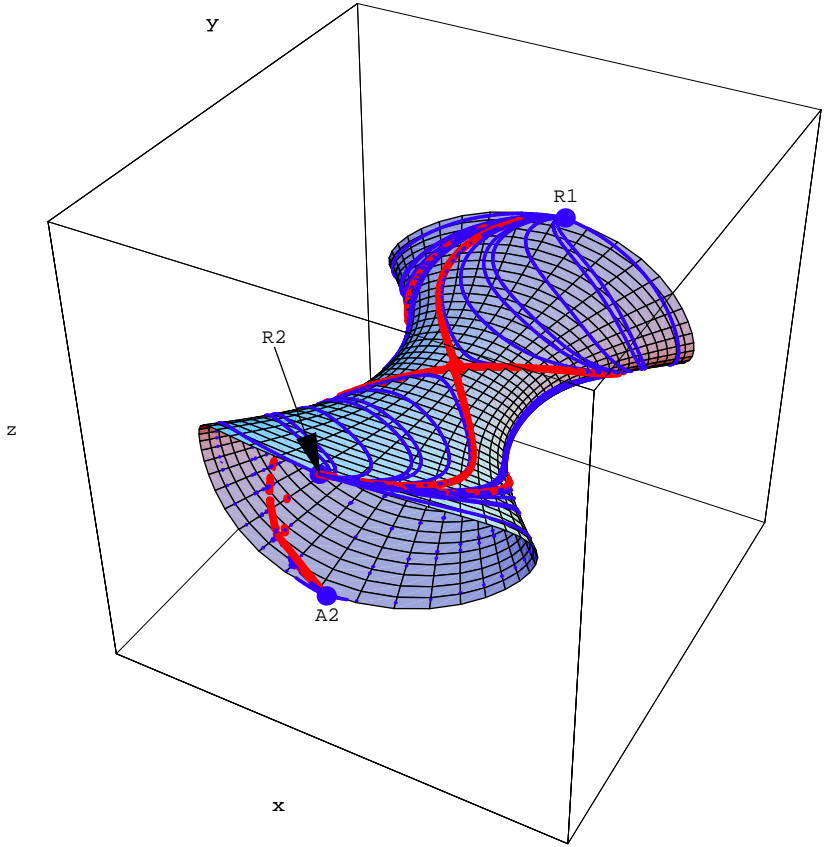},height=0.55\textheight}}
\vspace{-1.0em}
\caption{Trajectories with $H^2=0$ for $V=1/2 m^2 \phi^2 +
\Lambda$ with $\Lambda<0$.
}\label{m2x2lambdaminush0}
\end{figure}

Once again we take $m=1$ so $V = {1 \over 2} x^2 + \Lambda $ Since
$\Lambda$ is not divergent as $x \rightarrow \infty$ the set of 6
infinite critical points is the same as for $V=1/2 m^2 \phi^2$. To
consider the finite critical points and the limiting surface $H^2=0$,
however, we have to distinguish between two cases based on the sign of
$\Lambda$. The limiting surface is given by
\begin{equation}
y^2 = 12 z^2 + x^2 + 2 \Lambda,
\end{equation}
which is a hyperboloid of one or two sheets for $\Lambda<0$ and
$\Lambda>0$ respectively. For $\Lambda>0$ the potential is positive
definite and there are no finite critical points. For $\Lambda<0$
there are two finite saddle points at $x=y=0$, $z=\pm
\sqrt{\vert\Lambda\vert/6}$.

For $\Lambda>0$ the phase portrait is nearly identical to the case
studied above. The only difference is that the two cones separate into
two disconnected pieces of a hyperboloid. Since trajectories for
$\Lambda=0$ could not pass from one cone to the other anyway, this
change has no significant effect on the behavior of the system.

Figure~\ref{m2x2lambdaminush0} shows the phase portrait for the
quadratic potential with a negative cosmological constant. There are
no longer two separated sheets, reflecting the fact that there are now
trajectories connecting positive and negative $y$. There are two
critical saddle points at which the field is sitting motionless at the
top of the hill.  There are eight separatrix trajectories, one each
connecting each of the four infinite critical points to each of the
finite critical points on the limiting surface. As before all
trajectories begin and end at the $y$ dominated infinity.  If we fix
the brane at the middle of a trajectory, one of the singularity will
be screened. However, without the second brane it is impossible to
screen the other singularity. Therefore, the RS model with 
negative cosmological constant  and a
single $Z_2$ symmetric brane will acquire singularity at final distance
 if a massive scalar
field with nonvanishing condensate $\phi(w)$ is added.

\section{Exponential Bulk Potentials}

\begin{figure}

\begin{minipage}[t]{.45\columnwidth}
\centerline{\psfig{figure=\picdir{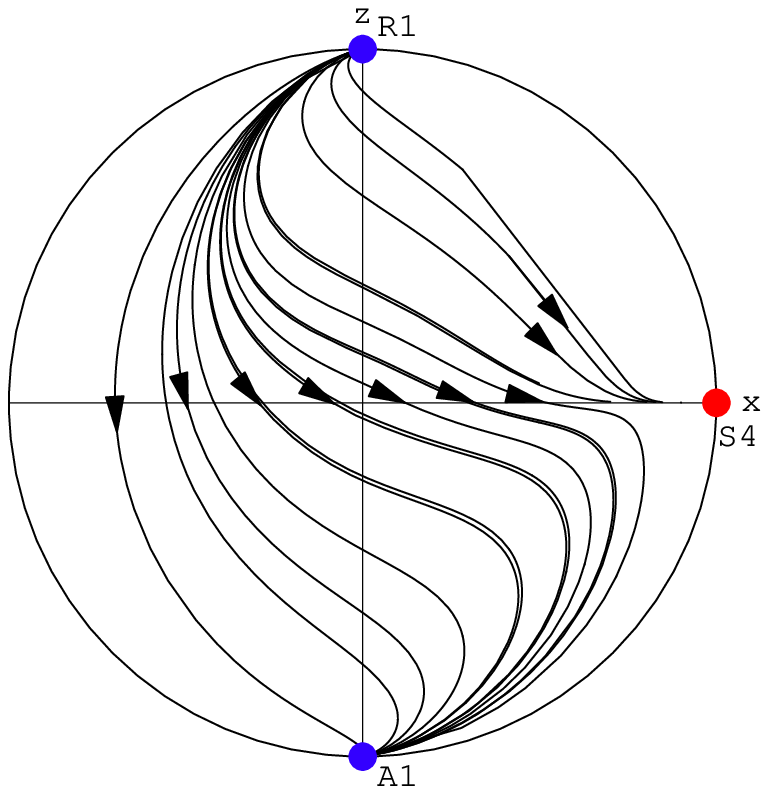},width=\columnwidth}}
\end{minipage}
\begin{minipage}[t]{.45\columnwidth}
\centerline{\psfig{figure=\picdir{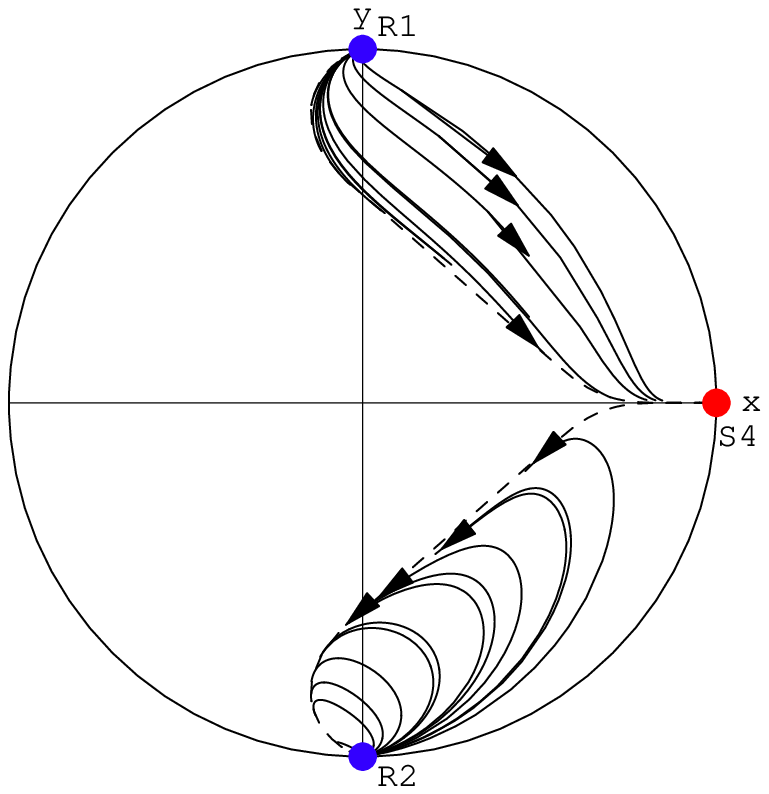},width=\columnwidth}}
\end{minipage}
\caption{2d projections of trajectories for $V=e^{-2 \sqrt{2} \phi}$,
$H^2=0$ on $xz, (y > 0)$ and $xy$ planes. In both projections, only
half of the surface is shown for clarity.}\label{exph02d}
\end{figure}

In this section we consider the bulk scalar potential $V =V_0 e^{-2
\sqrt{2} \phi}$. Dilaton scalar fields with exponential potentials
naturally arise in many high dimensional theories. 
 We consider the properties of warped geometry and dilaton using the
phase portrait of their dynamical system.  In our dimensionless
variables, after rescaling of $w$ by $V_0^{1/2}$, we have $V=e^{-2
\sqrt{2} x}$.

Since the potential is positive definite there are no finite critical
points. Infinite critical points will not occur at any point where the
potential diverges exponentially, i.e. for $x<0$. There are thus 7
infinite critical points 
\begin{eqnarray}\label{ipp}
(\varphi,\theta)
&=&(0,0), ~ (\pi/2,0), ~ (\pi,0), \\
&& (\sin^{-1}(\sqrt{12/13}), \pi/2), ~ (\sin^{-1}(\sqrt{12/13}), 3\pi/2), \nonumber\\
&& (\pi-\sin^{-1}(\sqrt{12/13}), \pi/2), ~ (\pi-\sin^{-1}(\sqrt{12/13}), 3\pi/2) \nonumber
\end{eqnarray}
corresponding to all points shown in Figure~\ref{fig:crit} except $S_3$.

The limiting surface $H^2=0$ is described by
\begin{equation}
y^2 = 12 z^2 + 2 e^{-2 \sqrt{2} x}.
\end{equation}
Like for quadratic case, this surface consists of two branches, one each
at positive and negative $y$, that touch only at a critical point. In
this case that critical point is at $x=\infty$.
For $H^2=0$ we once again have that $y$ can never change sign, meaning
a field moving up the potential from negative infinity will always
continue on to $x=+\infty$. There are, however, two very different ways
this can occur. Figure~\ref{exph02d} shows 2D projections of the phase
portrait for this case.  Consider upper half plane $y > 0$. Many
trajectories begin at infinite point $R_1$ $z>0$, $x/y \to 0$ and end at
infinite point $A_1$ $z<0$, $x/y \to 0$, just as they did for the
quadratic potential. For these trajectories $y$ is growing ever more
rapidly and a gradient singularity must occur at a finite distance
(from the fixed brane). There is, however, another class of
trajectories that approaches the infinite  critical point $S_4$ at
$z=y/x \to 0$. These are trajectories for which $y$ and $z$
asymptotically decrease as $x$ approaches infinity, so the field and
metric do not diverge in a finite distance. There is no singularity at
the end of trajectories which are heading towards $S_4$. Two types of
trajectories are divided by separatrix between $R_1$ and $S_4$.
Behaviour for $y < 0$ half is similar.

All trajectories in Figure~\ref{exph02d} for exponential bulk potential
can be obtained analytically as closed form solutions of
Hamilton-Jacobi equation (\ref{hjcon1}) derived in
Section~\ref{sec:HJ}. However, we will not give their explicit form
here, as the behavior of the solutions is adequately described by the
phase portrait already discussed.

\section{Warped Geometry with 4d de~Sitter Slices}\label{dsspace}

In this section we extend the phase space analysis to include non-flat
4d sections of the 5d warped geometry with constant 4d curvature $\sim
H^2$.  We consider the case of positive $H^2$ corresponding to
de~Sitter 4d geometry.  We continue to consider our major example, a
simple quadratic potential $V(\phi)={ 1 \over 2} m^2 \phi^2$, which we
began to investigate in Section \ref{sec:quadratic}.

\begin{figure}[t!]
\centerline{\psfig{figure=\picdir{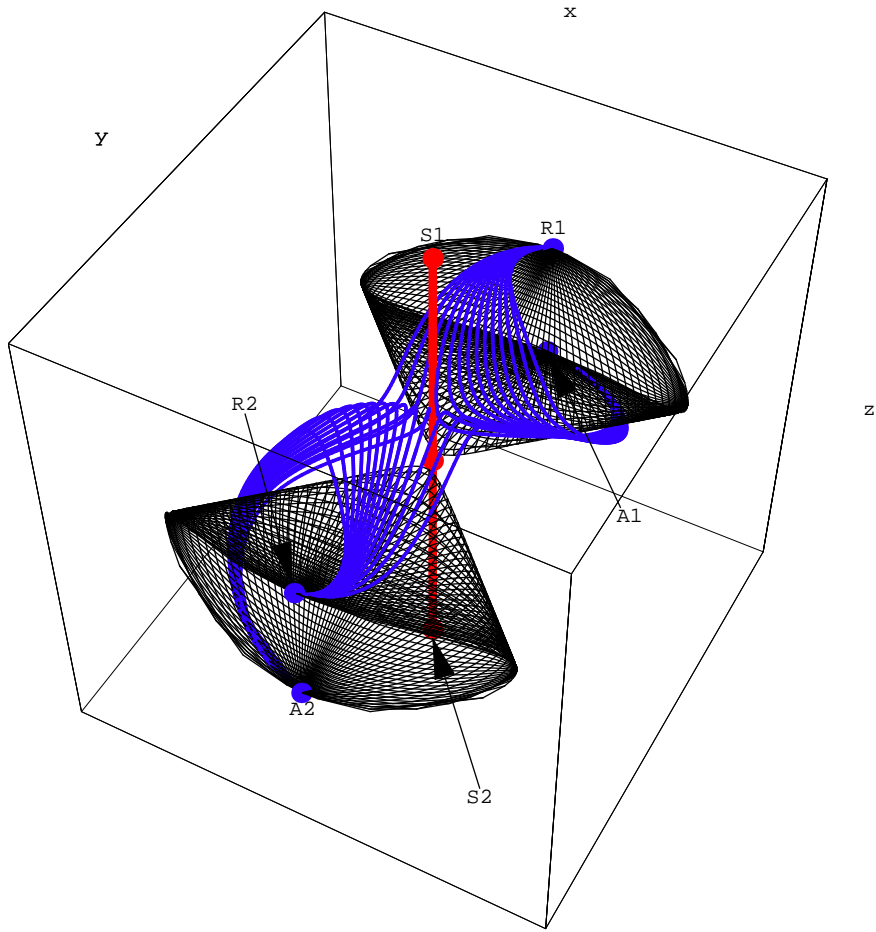},width=.75\columnwidth}}
\caption{Trajectories with $H^2>0$ for $V=\frac{1}{2} m^2 \phi^2$.}\label{m2x2}
\end{figure}

All trajectories in the 3d phase space $(x, y, z)$ in this case are
located outside the cone. Without branes, a sampling of trajectories
with $H^2>0$ is shown in Figure~\ref{m2x2}.

Once again, trajectories begin and end at the same four critical
points on the top and bottom edges of the cone. Now, however, there is
no topological constraint preventing $y$ from changing sign, so
trajectories beginning at either of the points at $z>0$ can end at
either of the points at $z<0$. As before, however, the warp factor
always goes from increasing to decreasing as the derivative $A'/A$
decreases monotonically from positive infinity to negative
infinity. There is once again a topological constraint imposed by a
pair of separatrices, however. In this case these are the trajectories
that move straight down along the line $x=y=0$, one between the point
$S_1$ and the origin $S$ and the other between the origin and the point
$S_2$. Trajectories passing to the right of this line must end at
$y>0$ and trajectories passing to the left of it must end at
$y<0$. This constraint simply reflects the fact that if the field
comes to rest on one side of the hill it will end up rolling down on
that same side. Once again all trajectories end up with $x/y \to 0$,
suggesting they reach the gradient type
singularity in a finite distance.

The most interesting feature of the flow of phase trajectories in
three dimensions is that they all tend to lean towards the limiting
surface of the cone in the region of positive $z$. The physical reason
for this is the following.  Inspecting the constraint equation
(\ref{constraint}) we see that for increasing $A(w)$ (positive $z={A'
\over A}$), the curvature term ${ H^2 \over A^2}$ decreases.  This
means that the 4d de~Sitter slices of the 5d geometry are getting
flatter.  This is very similar to how in 4d cosmology inflation
flattens the universe.  Imagine we fix two branes at $w_1$ and $w_2$,
and the warp factor is increasing between them, $A(w_2) > A(w_1)$. Let
us normalize $A(w_1)=1$.  Let us rewrite the 5d metric, specifying
the 4d de~Sitter coordinates in the form
\begin{equation}\label{desit}
ds^2=dw^2+A^2(w)\left(-dt^2+ e^{2Ht} d{\vec x}^2 \right) \ .
\end{equation}
The intrinsic 4d metric at the first brane at $w_1$ is
\begin{equation}\label{w1}
ds^2_4=\left(-dt^2+ e^{2Ht} d{\vec x}^2 \right) \ ,
\end{equation}
while the intrinsic metric at the second brane is 
\begin{equation}\label{w2}
ds^2_4= A^2(w_2) \left(-dt^2+ e^{2Ht} d{\vec x}^2 \right) \ .
\end{equation}
Rescaling the 4d coordinates with $A(w_2)$, $ t'=A(w_2) t$, ${\vec
x'}= A(w_2) {\vec x}$, we obtain 4d metrics in the canonical form
\begin{equation}\label{phys}
ds^2_4= -dt'^2+ e^{2H't'} d{\vec x}'^2 \ .
\end{equation}
The physical curvature is described in terms of $H'=H/A(w_2)$, which
is smaller than $H$, and thus the second brane is flatter than the
first one. The larger $A(w_2)$, the flatter the brane at $w_2$. It
would be interesting to apply this mechanism to the problem of
smallness of the cosmological constant on the visible brane.

\section{Phase Trajectories, Einstein-Hamilton-Jacobi\\
Equations, and the SUSY Form of the Potential}\label{sec:HJ}

So far we have only considered warped geometry with a bulk scalar
without including branes.  Branes must be self-consistently embedded
in the 5d spacetime in accordance with the junction conditions. The
junction conditions involve the brane scalar field potential
$U(\phi)$. Their solution is often found by using an (auxiliary) SUSY
superpotential \cite{Dewolfe}.

Before we apply our phase portrait methods to the brane world scenario
with branes, we introduce one more element. Phase space trajectories
can be conveniently described in terms of Hamilton or Hamilton-Jacobi
equations. In general relativity, usually the ADM $3+1$ formalism is
used to derive the Einstein-Hamilton-Jacobi equations. In the context
of 4d FRW cosmology with a scalar field the Einstein-Hamilton-Jacobi
equations were derived by Bond and Salopek \cite{BS90}. In the context
of holographic renormalization group flows, the Hamilton-Jacobi
equations were considered by de~Boer {\it et.~al.} \cite{deBoer:1999}.
In this section we extend these results for a $D$-dimensional spacetime
with a scalar field. We use $(D-1)+1$ splitting, but our $(D-1)$
hypersurface can be either timelike or spacelike, and we consider
branes of arbitrary constant curvature, and not just the flat case. We
find that for the flat brane geometry, the constraint equation reduces
to the SUSY representation of the scalar potential (this occurs even
though the system is not necessarily supersymmetric).

Let us consider an arbitrary D-dimensional metric in Gaussian normal
coordinates,
\begin{equation}
ds^2 = \epsilon\, dw^2 + g_{ab} dx^a dx^b \ ,
\end{equation}
where $\epsilon=\pm1$ depending on the timelike or spacelike character
of the $(D-1)+1$ splitting. For the 4d cosmological problem
\cite{BS90} $\epsilon=-1$, while for the 5d warped geometry
$\epsilon=+1$.

The components of the curvature tensor can be split according to the
Gauss-Codazzi equations
\begin{eqnarray}
{^{(D)}}R^{a}_{~bcd} &=& {^{(D-1)}}R^{a}_{~bcd} +
\epsilon(\K^{a}_{~d}\K_{bc} - \K^{a}_{~c}\K_{bd})\nonumber\\
{^{(D)}}R^{w}_{~abc} &=& \epsilon(\K_{ab:c} - \K_{ac:b})\nonumber\\
{^{(D)}}R^{w}_{~awb} &=& \epsilon(-\K_{ab,w} + \K_{ac}\K^{c}_{~b})
\end{eqnarray}
The Einstein and Ricci tensor components can be split as
\begin{eqnarray}
  {^{(D)}}G^{w}_{~w} &=& - \frac{1}{2}\, {^{(D-1)}}R
     + \frac{\epsilon}{2}(\K^2 - \K_{ab}\K^{ab})\nonumber\\
  {^{(D)}}G^{w}_{~a} &=& \epsilon(\K^{~c~}_{a~:c} - \K_{:a})\nonumber\\
  {^{(D)}}R_{ab} &=& {^{(D-1)}}R_{ab}
     + \epsilon(2\K_{a}^{~c}\K_{bc} - \K\K_{ab} -\K_{ab,w}) \ .
\end{eqnarray}

We further specify the metric ansatz in a form unifying equations
(\ref{warp}) and (\ref{fr})
\begin{equation}
  ds^2 = \epsilon\, dw^2 + A^2(w) \gamma_{ab}(x^i) dx^a dx^b \ ,
\end{equation}
where $ \gamma_{ab}$ is the metric of $D-1$ dimensional constant
curvature space, and we write
\begin{equation}\label{hamilt}
  \H = \frac{A'}{A} \ .
\end{equation}
Although this function of the warp factor was already used for the
``Hubble'' parameter $z$, here we have denoted it differently since the
very same combination can be considered as a function of $\phi$ and 
will play the role of the Hamiltonian.

For this metric ansatz, we have
\begin{eqnarray}
  &
  \K_{ab} = \frac{1}{2}\, g_{ab,w} = \H g_{ab}, \hspace{1em}
  \K = (D-1)\H,
  &\\&
  \K_{ab}\K^{ab} = (D-1)\H^2, \hspace{1em}
  \K_{ab,w} = (\H'+2\H^2) g_{ab}.
  \nonumber
\end{eqnarray}

The bulk Einstein equations give us three equations. The first
equation is
\begin{equation}\label{mom}
  - (D-2) \H_{,a} = \phi' \phi_{,a} \ .
\end{equation}
It can be shown \cite{BS90} that $\H=\H(\phi)$. Then from (\ref{mom})
we obtain the momentum constraint equation
\begin{equation}
  - (D-2) \H_{,\phi} = \phi'  \ .
\end{equation}

The second equation is the energy constraint equation
\begin{equation}\label{dyn}
  {\textstyle \frac{1}{2}(D-1)(D-2)} \H^2 - 
 {^{(D-1)}}R \, \frac{\epsilon}{2}    =
    \frac{\phi'^2}{2} - 
\epsilon \left[\textstyle\frac{1}{2} \phi_{,a}\phi^{,a} +V(\phi)\right] \ .
\end{equation}

The third equation is 
\begin{equation}\label{dyn1}
  {^{(D-1)}}R_{ab} - \epsilon\{(D-1)\H^2 + \H'\} g_{ab} =
    \phi_{,a}\phi_{,b} + {\textstyle \frac{2}{D-2}\,} V(\phi)\, g_{ab}
\end{equation}
From the trace of the last equation and the energy constraint it
follows that
\begin{equation}\label{dyn2}
  {\textstyle \frac{D-1}{2}} (\H'+\H^2) + \frac{\phi'^2}{2} + {\textstyle
\frac{\epsilon}{D-2}} V(\phi) = 0.
\end{equation}

Next we suppose that $\phi=\phi(w)$. Then from (\ref{dyn}--\ref{dyn2})
we find that the scalar field configuration in D dimensions is
described by a system of three variables $\{\phi,\phi',\H\}$
\begin{eqnarray}
  \H' &=& -\H^2 - \frac{\phi'^2}{D-1} - \frac{2\epsilon}{(D-1)(D-2)}\, V(\phi) \ ,
\nonumber\\
  \phi'' &=& \frac{\partial V}{\partial\phi} - (D-1)\H \phi' \ .
\end{eqnarray}
These equations generalize the equations
(\ref{einstein1}--\ref{phiequation}) and
(\ref{einstein3}--\ref{phiequation1}).

So far we have not specified the curvature of the $D-1$ hypersurfaces.
Assuming that the induced brane-world metric is de~Sitter spacetime,
${^{(D-1)}}R_{ab}=(D-2)\frac{H^2}{A^2}\,\gamma_{ab}$, and from
(\ref{dyn}) we obtain the constraint equation
\begin{equation}\label{hjcon}
\epsilon V(\phi) =
  \frac{1}{2} (D-2)^2 \left(\frac{\partial\H}{\partial\phi}\right)^2
- \frac{1}{2} (D-1)(D-2) \left(\H^2 - \epsilon \frac{H^2}{A^2}\right) \ .
\end{equation}

Let us apply the Hamilton-Jacobi constraint equation for warped
geometry with $\epsilon=+1$ and flat $D-1$ dimensional slicing
with $H^2=0$.  In
this case we have
\begin{equation}\label{hjcon1}
 V(\phi) =
  \frac{1}{2} (D-2)^2 \left(\frac{\partial\H}{\partial\phi}\right)^2
- \frac{1}{2} (D-1)(D-2) \H^2 \ .
\end{equation}
Here the scalar field potential is taken as function of $\phi$
which is solution of the self-consistent equations (in loose terminology,
``on-shell'' value of the potential)
It is not expected that the arbitrary potential $V(\phi)$
will have the SUSY form (\ref{hjcon1}) for non self-consistent
geometry.

Compare this equation with the well known result that the stability of
the bulk scalar field requires a ``supersymmetric'' form of the
potential \cite{susy1,susy2}
\begin{equation}\label{susy}
 V(\phi) = 2(D-2)^2 \left(\frac{\partial W}{\partial\phi}\right)^2 - 
 2(D-1)(D-2) W^2 \ ,
\end{equation}
where $W$ is some auxiliary function $W=W(\phi)$ which is called the
``superpotential'', but which emerges from the requirement of
stability even without supersymmetry.

Comparing equations (\ref{hjcon1}) and (\ref{susy}), we see that the
bulk potential (\ref{hjcon1}) can be expressed in the SUSY form
(\ref{susy}) where the Hamiltonian $\H$ plays the role of the
superpotential $W=\frac{1}{2}\H$.  It would be interesting to
understand the relation between these two apparently different
approaches to scalar field dynamics in D dimensional spacetime that
lead to such similar looking results. Here we have simply noted that
the SUSY form of the potential is a very convenient to treat the bulk
and the junction conditions in a unified way.

The brane junction conditions (\ref{junction1},\ref{junction2}) in
terms of the Hamiltonian $\H$ are
\begin{equation}
  -2(D-2) \H = U(\phi), \hspace{1em}
  -2(D-2) \frac{\partial\H}{\partial\phi} = \frac{\partial U}{\partial\phi} \ ,
\end{equation}
where the values of all the functions are taken at the brane position
$w$.  Therefore, choosing $U(\phi) = -2(D-2)\H(\phi)$ at the brane, we
automatically satisfy the junction conditions.  This is similar to the
approach of \cite{Dewolfe}. Here, however, we make an explicit
connection between the superpotential $W$ and the Hamiltonian $\H$.

We also shall notice that for curved branes with nonzero curvature
$H^2$, the form (\ref{hjcon1}) is not supersymmetric anymore.

\section{Warp Geometry Between Branes}

In the previous sections we discussed how to solve the gravity/scalar field
equations in the bulk. To complete the brane world picture we must
include branes as well, which are embedded in the bulk according to
junction conditions (\ref{junction1},\ref{junction2}). We will consider
the case of static branes ($w_i = \text{const}$) only, and assume mirror
symmetry across the brane ($Z_2$ symmetry).

\begin{figure}[t!]
  
  
  \centerline{\psfig{figure=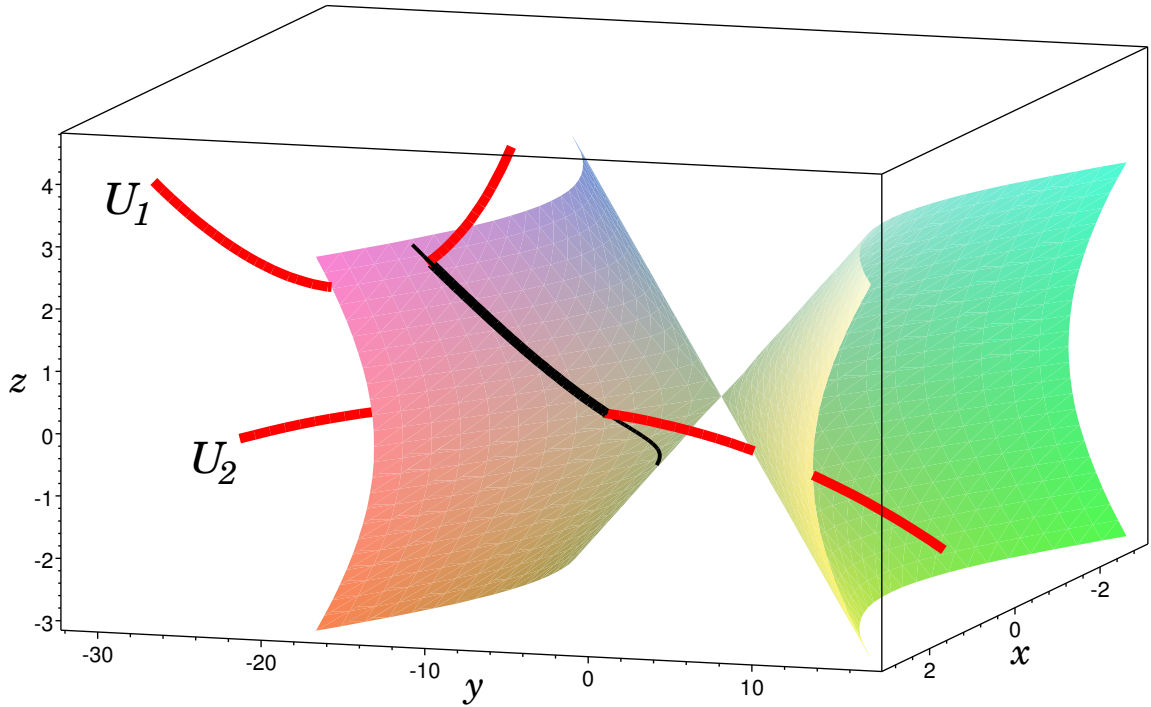,height=3.75in}}
  \caption{
    Segment of trajectory at the cone (shaded surfaces) in $\frac{1}{2}m^2 \phi^2$ theory
    between two branes. Phase space positions of branes are obtained by
    intersection of the trajectory with  $U_1,U_2$ curves, and is
    ``pinned'' by the choice of brane surface potentials $U_1,U_2$.
  }\label{pinning}
\end{figure}

In the language of phase portraits, the two boundary conditions
(\ref{junction1},\ref{junction2}) define a 1d curve in the 3d phase
space. The equation for this curve is determined by the surface
potential $U(\phi)$, and can be parameterized by the value of the scalar
field $\phi$ as
\begin{equation} \label{ucurve}
  y = \pm \frac{1}{2}\, U_{,x}(x), \hspace{1em}
  z = \mp \frac{U(x)}{6},
\end{equation}
where the signs  correspond to the orientation  (the choice of normal
vector $n_{\mu}$) of the brane with respect to the bulk: the
upper/lower signs are taken for the brane being on the left/right edge
of the bulk, i.e. at lower/higher limit of $w$. The bulk trajectories
must start/end on these curves in the full phase space (and not at
infinity as it was in the absence of branes) for junction conditions to
be satisfied at the brane, as illustrated in Figure~\ref{pinning}. The
location of the brane is then given by the point of intersection of the
phase space curve (\ref{ucurve}) and the bulk trajectory, and can be
uniquely specified by the value of the field $\phi$ on the brane.

If one considers the case of a brane world with a single brane,  for
many potentials $V(\phi)$ one encounters a problem that, in general,
singularities occur a finite distance from the brane (although  there
are potentials like exponential potential considered in Section 6,
where singularity outside of $Z_2$ symmetric brane is avoidable). One
possible solution to the  problem of timelike singularity outside of
the brane  in the brane world scenario is to shield the singularity
with a second brane. Given our assumption of $Z_2$ symmetry, the
imposition of a second brane would effectively compactify the fifth
dimension, making a circle with the two branes at opposite poles. The
junction conditions must then be met separately at each brane, which is
not necessarily guaranteed for an arbitrary choice of potentials $V$
and $U$.

Let us address the question about the existence of warped geometry
bulk  solution between two branes, if the potentials $U_i$ at the brane
are arbitrary. It is known \cite{GW,Dewolfe} that in general such
solutions can exist, but they are not expected to be compatible with
the flat brane, but rather with curved brane. We can obtain similar
conclusion using the methods developed in this paper. To see this in
terms of phase portraits, the problem of finding a consistent two-brane
solution reduces to finding a trajectory in the full three dimensional
phase space  that connects the two corresponding junction condition
curves $U_1$ and $U_2$. Geometrically, this can be viewed as finding
the intersection of a 2d surface generated by the bulk trajectories
passing through the first curve $U_1$, and the second curve $U_2$.

As an intersection of a 2d surface and a curve in 3d is typically a
point (unless they overlap or do not intersect altogether), for generic
choice of $V$ and $U$ there may be only one such trajectory (if any),
and the brane separation is fixed by its length. However, this
successful trajectory is not expected in general to be located on the
$H^2=0$ surface. Still, one can force the segment of trajectory between
branes to be on the surface $H^2=0$ by simply shifting the potentials
$U_1$, $U_2$ by a constant amount. This tuning is a familiar tuning of
four dimensional cosmological constant on the brane (see also
\cite{Dewolfe}).

We conclude that for potentials  $U_i$ which are not specially
selected, brane worlds scenario is not expected to have flat branes.
The branes in the self-consistent scenario are flat in a special but
important case, which  occurs when the junction condition curve is also
a bulk trajectory, as it happens for Hamiltonian of the previous
section. In this case, the two branes of opposite potentials $U_1 =
-U_2 = -2(D-2) \H$ can consistently be placed at any separation. If one
modifies the brane surface potentials $U_{1,2}$ by adding quadratic
corrections $\propto (\phi-\phi_{1,2})^2$ as it was done in
\cite{Dewolfe}, the solution will be ``pinned'' between $\phi_1$ and
$\phi_2$, as illustrated in Figure~\ref{pinning}, and the brane
separation will be fixed by the choice of $\phi_1$ and $\phi_2$. In
other words, the separation in the space of $\phi$ is translated into
the inter-brane distance.

\section{Conclusion}

We have developed here a method for systematically exploring the
properties of different potentials in brane world  warped geometry. To
construct the  phase portrait of the dynamical system of
gravity/scalar, one can apply the qualitative theory of differential
equations. Solutions of these equations are represented by the
trajectories propagating in the phase space. For a single bulk scalar,
trajectories are in three dimensional  phase space. For the case of the
flat branes, all trajectories are located at two dimensional surface,
and the phase portrait of the dynamical system can be easily
investigated.

In general, the phase space trajectories have timelike singularities at
one or two of their ends. These singularities are dominated by the
scalar field gradient term, and associated with the infinite critical
points in the phase space. We describe how to find critical points for
an arbitrary potentials. There are, however, examples of the potentials
without singularity at one of the end of phase trajectory. In this case,
it is possible to construct non-singular warped geometry with a single
brane with $Z_2$ symmetry.

We also considered the Einstein-Hamilton-Jacobi formulation of the 
warped geometry with scalar field.  Constrain equation relate the
arbitrary bulk potential, the Hamiltonian and its $\phi$-derivative.
Surprisingly, the scalar potential taken on the self-consistent
solutions acquires SUSY form even without underlying supersymmetry in
the theory. We address the issue how this  form of the constrain
equation for arbitrary ``on-shell'' potential is related to the
requirement to the of the SUSY form of the potential for gravitational
stability of gravity/scalar system.

One can use the phase space with bulk trajectories to study  warped
geometry between two branes. Junction conditions for each brane
generate  one dimensional curve in the phase space. Segment of
trajectory between two such curves  corresponds to the inter-brane warp
factor and scalar field. Without tuning the potential, this
configuration in general is not located at the two dimensional surface
which represents the flat branes, in other words, the solution exist in
general for curved branes. However, one can achieve solution with two
flat branes by a simple shift of the potentials. This analysis can be
easily extended for more realistic case of several bulk scalar degrees
of freedom. For instance, for two bulk scalars, phase space is five
dimensional and brane junction conditions generate two dimensional
surface. Without tuning the brane potentials, in general there is a
segment of trajectory which connect both two dimensional surfaces in
5d (except special cases).

We leave the phenomenological applications of our methods for
construction of braneworld scenario with stabilization and
investigation of their stability for future work.

\section*{Acknowledgements}

We are grateful to Dick Bond, Andrei Linde, Renata Kallosh, Boris
Khesin, and Dario Martelli for fruitful discussions and comments.
This work was supported by NSERC, CIAR, PREA of Ontario and NATO
Linkage Grant 975389.

\appendix
\section*{Appendix}

In this appendix we find the coordinates of the infinite critical
points using the coordinates of the Poincar\'{e} mapping. We continue
to denote derivatives with primes, with the understanding that all $r$,
$\theta$, and $\varphi$ derivatives are with respect to $\tilde{w}$. In
these new variables, our basic equations 
(\ref{phaseequation1}--\ref{phaseequation3}) for arbitrary potential
become
\begin{eqnarray}
  r' &=&
    {1 \over 16}\, (1-r) \cos\varphi
      \left(-25 + 9 \cos 2\varphi + 34 \cos 2\theta \, \sin^2\varphi\right)
\\ \nonumber &&
  + {1 \over 8}\, (1-r)^2
      \left(4 \sin 2\theta \, \sin^2\varphi - \cos\varphi
      (-25 + 9 \cos 2\varphi + 34 \cos 2\theta \, \sin^2\varphi)\right)
\\ \nonumber &&
  + {1 \over 48}\, (1-r)^3
      \left(24 \sin\varphi (-2 V_{,x} \sin\theta + \sin 2\theta \, \sin\varphi)
\right. \\ \nonumber && \hspace{6em} \left.
          + \cos\varphi (8 V + 75 - 27 \cos 2\varphi - 102 \cos 2\theta \, \sin^2\varphi)\right),
\\ \theta' &=&
    - 2r \cos\varphi \, \sin 2\theta
    - (1-r) \sin^2\theta
    + \frac{(1-r)^2}{r}\, V_{,x} \cos\theta \, \csc\varphi,
\\ \varphi' &=&
    {1 \over 32}\, r
      \left(- 14 \sin\varphi - 18 \cos 2\varphi \, \sin\varphi
            + 13 \cos 2\theta \, \sin\varphi + 17 \cos 2\theta \, \sin 3\varphi\right)
\\ \nonumber &&
  + {1 \over 4}\, (1-r) \sin 2\theta \, \sin 2\varphi
  + \frac{(1-r)^2}{r}\, \left(V_{,x} \cos\varphi \, \sin\theta + {1 \over 6}\, V \sin\varphi\right).
\end{eqnarray}

Suppose that the potential $V(x)$ and its derivative $V_{,x}$ are not
divergent at infinity $x \to \infty$, and there is no ``accidental''
cancellations of terms (see example below). Taking limit $r=1$ and 
putting {\it r.h.s.} of equations for $\theta'$, $ \varphi'$ to zero,
we find 8 solutions for $\theta, \varphi$ at the Poincar\'{e} sphere,
given by equation  (\ref{infcrit}).

For quadratic potential $V(x)=\frac{1}{2}x^2$.  we plug the expressions
for $V$ and $V_{,x}$ into the above dynamical equations. The behavior
of $r'$ and $\theta'$ at $r=1$ are unchanged. The equation for
$\varphi'$, however, picks up another nonzero term in that limit,
namely
\begin{equation}
  {1 \over 6}\, \frac{(1-r)^2}{r}\, V \sin\varphi =
  {1 \over 12}\, \cos^2\theta \, \sin^3\varphi,
\end{equation}
so the total $\varphi'$ equation becomes
\begin{equation}
  \varphi' = {1 \over 96}\, \left\{
    (- 39  - 54 \cos 2\varphi + 42 \cos 2\theta) \sin\varphi
    + (50 \cos 2\theta - 1) \sin 3\varphi
  \right\}.
\end{equation}
The {\it r.h.s.} of  equations for  $\theta'$, $ \varphi'$
vanish for 6 points $(\theta, \varphi)$, given by equation (\ref{ip}).
Two points from (\ref{infcrit}) disappear due to the cancellation by
special form of the potential $V$.

For exponential potential we have 
\begin{equation}
  V = e^{-2\sqrt{2} x}
    = \exp\left(-2\sqrt{2}\, {r \over 1-r}\, \cos\theta \, \sin\varphi \right),
\end{equation}
\begin{equation}
V_{,x} = -2\sqrt{2} e^{-2\sqrt{2} x}
    = -2\sqrt{2}\, \exp\left(-2\sqrt{2}\, {r \over 1-r}\, \cos\theta \, \sin\varphi \right).
\end{equation}
At infinite point with $x < 0$ potential diverges exponentially,
and one of the 8 critical points  (\ref{infcrit}) disappear.
Remaining 7 points are given by equation (\ref{ipp}).


\begin{thebibliography}{999}

\bibitem{HW}
P.~Horava and E.~Witten,
{\it Heterotic and type I string dynamics from eleven dimensions},
Nucl. Phys. {\bf B460}, 506-524 (1996) [arXiv:hep-th/9510209].

\bibitem{Lukas}
A.~Lukas, B.~A.~Ovrut, K.~S.~Stelle and D.~Waldram,
{\it Heterotic M-theory in five dimensions},
Nucl. Phys. {\bf B552}, 246-290 (1999) [arXiv:hep-th/9806051].

\bibitem{RS1}
L.~Randall and R.~Sundrum,
{\it A large mass hierarchy from a small extra dimension},
Phys. Rev. Lett. {\bf 83}, 3370-3373 (1999) [arXiv:hep-ph/9905221].

\bibitem{RS2}
L.~Randall and R.~Sundrum,
{\it An alternative to compactification},
Phys. Rev. Lett. {\bf 83}, 4690-4693 (1999) [arXiv:hep-th/9906064].

\bibitem{GW}
W.~Goldberger and M.~Wise,
{\it Modulus stabilization with bulk fields},
Phys. Rev. Lett. {\bf 83} 4922-4925, (1999) [arXiv:hep-ph/9907447].

\bibitem{Dewolfe}
O.~DeWolfe, D.~Freedman, S.~Gubser and A.~Karch,
{\it Modeling the fifth dimension with scalars and gravity},
Phys. Rev. {\bf D62}, 046008 (2000) [arXiv:hep-th/9909134].

\bibitem{GKL}
G.~Gibbons, R.~Kallosh and A.~Linde,
{\it Brane world sum rules},
JHEP {\bf 0101}, 022 (2001) [arXiv:hep-th/0011225].

\bibitem{AVP}
A.~Van~Proeyen,
{\it The scalars of $N=2$, $D=5$ and attractor equations},
arXiv:hep-th/0105158.

\bibitem{CR}
H.~Chamblin and H.~Reall,
{\it Dynamic dilatonic domain walls},
Nucl. Phys. {\bf B562}, 133-157 (1999) [arXiv:hep-th/9903225].

\bibitem{ST}
K.~Skenderis and P.~Townsend,
{\it Gravitational stability and renormalization-group flow},
Phys. Lett. {\bf B468} 46-51, (1999) [arXiv:hep-th/9909070].

\bibitem{deBoer:1999}
J.~de Boer, E.~Verlinde and H.~Verlinde,
{\it On the holographic renormalization group},
JHEP {\bf 0008}, 003 (2000)
[arXiv:hep-th/9912012].

\bibitem{FTW}
E.~Flanagan, S.-H.~Tye and I.~Wasserman,
{\it Brane world models with bulk scalar fields},
Phys. Lett. {\bf B522}, 155-165 (2001) [arXiv:hep-th/0110070].

\bibitem{Davis}
S.~Davis,
{\it Brane cosmology solutions with bulk scalar fields},
arXiv:hep-ph/0111351.

\bibitem{KLS}
L.~A.~Kofman, A.~D.~Linde, A.~A.~Starobinsky,
{\it Inflationary Universe generated by the combined action of a scalar field and gravitational vacuum polarization},
Phys.Lett. {\bf B157}, 361-367 (1985).

\bibitem{acad}
V.~A.~Belinskii, L.~P.~Grishchuk, Ya.~B.~Zel'dovich, and I.~M.~Khalatnikov,
{\it Inflationary stages in cosmological models with a scalar field},
Zh. Eksp. Teor. Fiz. {\bf 89}, 346-360 (1985).

\bibitem{fastroll}
A.~Linde,
{\it Fast-roll inflation},
JHEP {\bf 0111}, 052 (2001) [arXiv:hep-th/0110195].

\bibitem{FFKL}
G.~Felder, A.~Frolov, L.~Kofman and A.~Linde,
{\it Cosmology with negative potentials},
arXiv:hep-th/0202017.

\bibitem{BS90}
D.~S.~Salopek and J.~R.~Bond,
{\it Nonlinear evolution of long wavelength metric fluctuations in inflationary models},
Phys. Rev. {\bf D42}, 3936-3962 (1990).

\bibitem{susy1}
W.~Boucher,
{\it Positive energy without supersymmetry},
Nucl. Phys. {\bf B242}, 282 (1984).
 
\bibitem{susy2}
P.~Townsend,
{\it Positive energy and the scalar potential in higher dimensional (super)gravity theories},
Phys. Lett. {\bf B148}, 55 (1984).

\end{thebibliography}
\end{document}